\documentclass[preprint,showpacs,preprintnumbers,amsmath,amssymb,tightenlines,nofootinbib]{revtex4-1}
 \usepackage{graphicx}
 \usepackage{dcolumn}
 \usepackage{bm}
 \usepackage{ulem}
 \usepackage{enumitem}
 \usepackage{slashed}
 \usepackage{amssymb}
\def\q{\bm{q}}
\def\p{\bm{p}}
\def\k{\bm{k}}
\def\v{\bm{v}}
\def\x{\bm{x}}

\def\vec\epsilon{\bm{\epsilon}}

\def\q{\bm{q}}
\def\p{\bm{p}}
\def\k{\bm{k}}
\def\v{\bm{v}}
\def\x{\bm{x}}

\usepackage{xcolor}

\begin{document}

\title{Heavy quark diffusion and radiation at intermediate momentum}
\author{Juhee Hong}
\affiliation{Department of Physics and Institute of Physics and Applied Physics, Yonsei University,
Seoul 03722, Korea}
\date{\today}

\begin{abstract}
We discuss heavy quark diffusion and radiation in an intermediate-momentum 
regime where finite mass effects can be significant. 
Diffusion processes are described in the Fokker-Planck approximation 
for soft momentum transfer, while radiative ones are taken into account by 
nearly collinear gluon emission from a single scattering in the Boltzmann 
equation.  
We also consider radiative corrections to the transverse momentum 
diffusion coefficient, which are $\mathcal{O}(g^2)$ suppressed than the 
leading-order diffusion coefficient but logarithmically enhanced. 
Numerical results show that the heavy quark distribution function depends on 
the energy loss mechanism so that the medium modifications by diffusion and 
radiation are distinguishable. 
Employing the heavy quark diffusion coefficient constrained by lattice QCD 
data, we estimate the nuclear modification factor which exhibits a 
transition from diffusion at low momentum to radiation at high momentum. 
The significance of the radiative effects at intermediate momentum depends on  
the diffusion coefficient and the running coupling constant. 
\end{abstract}

\maketitle

\section{Introduction}

Heavy quarks are important probes for high-temperature QCD matter created in 
relativistic heavy-ion collisions, as they are mostly produced at an early 
stage and conserved during the evolution. 
Slowly moving heavy quarks experience a Brownian motion in quark-gluon plasmas, 
and gluon-bremsstrahlung can affect the high-momentum spectra. 
Medium modifications of heavy quark production can be described by the 
collisional and radiative energy loss. 
Heavy quark transport and the related energy loss have been 
thoroughly investigated by various models (for recent review, see Refs. 
\cite{Dong:2019unq,Dong:2019byy,He:2022ywp}). 
Many of the transport models treat medium-induced gluon emission as 
an additional contribution to heavy quark diffusion or analogously to jet 
quenching with multiple scatterings. 
A recoil force term due to gluon radiation has been introduced 
in the Langevin equation for Brownian motion \cite{Cao:2013ita}, and the 
radiative energy loss has been estimated independently of the collisional 
energy loss \cite{Djordjevic:2003zk,Mustafa:2004dr}. 
In these previous studies, it is not easy to distinguish two 
energy-loss effects and to find out which mechanism is more influential, 
depending on momentum. 
This work introduces a heavy-quark transport approach that allows us to treat 
gluon-bremsstrahlung differently from diffusion while describing two 
mechanisms consistently with a single transport parameter. 
We concentrate on an intermediate-momentum regime where heavy 
mass effects can be significant and investigate the transition between 
diffusion and radiation from a single scattering.

The interaction between heavy quarks and dynamic thermal media is 
characterized by transport coefficients. 
Especially, the heavy quark diffusion coefficient depending on momentum and 
temperature is important because it controls the rate of 
equilibration in high-temperature QCD plasmas. 
The leading-order momentum diffusion coefficient has been calculated by 
hard-thermal-loop (HTL) perturbation theory 
\cite{Braaten:1991jj,Braaten:1991we,Moore:2004tg}, and its $\mathcal{O}(g)$ 
correction has been obtained in the soft sector \cite{Caron-Huot:2007rwy}. 
For a realistic value of the strong coupling constant, the classical 
correction is so large that nonperturbative determination is required. 
Similar to the jet transport parameter $\hat{q}$, there are also quantum 
corrections which are suppressed by $\mathcal{O}(g^2)$ but 
double-logarithmically enhanced \cite{Liou:2013qya,Blaizot:2013vha}. 
Recently, a Bayesian analysis and transport model comparison have been 
performed to determine the heavy quark transport coefficients from 
phenomenological studies \cite{Xu:2017obm,Rapp:2018qla,Cao:2018ews}. 
While most models are able to describe experimental data with some 
adjustment of parameters, the extracted diffusion coefficients 
vary due to the large differences between models.

The distribution function of heavy quarks can be described by the Boltzmann 
equation 
\begin{equation}
\label{boltz}
\left(\frac{\partial}{\partial t}+\v\cdot\frac{\partial}{\partial \x}\right)
f(\p)=C_{\rm col}[f]+C_{\rm rad}[f] \, ,
\end{equation}
where the collision terms correspond to elastic scattering and gluon 
emission for the collisional and radiative energy loss, respectively. 
In a leading-log approximation the first term can be formulated as a 
Fokker-Planck operator, while the second term is radiative corrections to the 
collision kernel responsible for diffusion. 
For heavy quarks with intermediate momentum, we can formulate the transport 
equation only in terms of the momentum diffusion coefficient which can be 
constrained by lattice QCD computations. 
With the single transport parameter, we can treat two types of energy loss 
consistently and study the relative importance of each mechanism in the 
transition region.

The outline of the paper is as follows. 
First, we briefly review the leading-log heavy quark diffusion with a 
Fokker-Planck equation in Section \ref{collision}. 
Then, we discuss the radiative effects, nearly collinear gluon-emission 
and radiative corrections to the transverse momentum diffusion coefficient 
in Section \ref{radiation}. 
In Section \ref{numerical}, we present the numerical results for the medium 
modifications of the heavy quark spectrum. 
Employing the heavy quark diffusion coefficient constrained by lattice 
QCD data and the running coupling constant, we estimate the 
nuclear modification factor of heavy quarks for a Bjorken expansion.  
In Section \ref{summary}, we summarize our results. 
The details on gluon emission are given in Appendix \ref{emit}.

\section{Heavy Quark Diffusion}
\label{collision}

We begin with a brief review on the collisional energy loss of heavy quarks 
in a relatively low-momentum regime 
\cite{Svetitsky:1987gq,Braaten:1991jj,Braaten:1991we,vanHees:2004gq,Moore:2004tg}. 
Traversing quark-gluon plasmas, heavy quarks with $m,p\gg T$ undergo diffusion 
by elastic scattering.
For spacelike soft-gluon exchange, the leading collision term in Eq. 
(\ref{boltz}) can be approximated as a Fokker-Planck operator,  
\begin{equation}
\label{Ccol}
C_{\rm col}[f]=\frac{\partial}{\partial p^i}\left[\eta(\p)p^if(\p)\right]
+\frac{1}{2}\frac{\partial^2}{\partial p^i\partial p^j}
\left[\kappa^{ij}(\p)f(\p)\right] \, ,
\end{equation}
where $\eta(\p)$ is the drag coefficient and 
$\kappa^{ij}(\p)=\kappa_L(p)\hat{p}^i\hat{p}^{j}+\kappa_T(p)(\delta^{ij}-\hat{p}^i\hat{p}^{j})$ is the momentum diffusion tensor.

For a heavy quark moving in the $z$-direction, the longitudinal and 
transverse momentum diffusion coefficients are defined as   
\begin{eqnarray}
\label{kappa}
\kappa_L(p)&=&\int d^3\q\frac{d\Gamma(\q)}{d^3\q}\q_z^2 \, ,
\nonumber\\
\kappa_T(p)&=&\frac{1}{2}\int d^3\q\frac{d\Gamma(\q)}{d^3\q}\q_T^2 \, ,
\end{eqnarray}
where $\q$ is the soft momentum transfer. 
Because the heavy quark mass is larger than a typical parton momentum 
of $\mathcal{O}(T)$, the dominant contribution comes from t-channel gluon 
exchange. 
In the Coulomb gauge, the collision rate is given by  
\begin{eqnarray}
\label{kernel}
C(\q)&\equiv&(2\pi)^3\frac{d\Gamma(\q)}{d^3\q} \, ,
\nonumber\\
&=&\frac{\pi}{2} g^2C_Fm_D^2
\int d\omega \, \delta(\omega-\q\cdot\v)\frac{T}{q}
\left[\frac{2}{|\q^2+\Pi_L(Q)|^2}+\frac{(q^2-\omega^2)(q^2v^2-\omega^2)}
{q^4|\q^2-\omega^2+\Pi_T(Q)|^2}\right] \, ,
\end{eqnarray}
where the interaction rate can be expressed in terms of the imaginary part 
of the heavy quark self-energy \cite{Braaten:1991jj}. 
Taking account of heavy quark interactions with both gluons and light quarks, 
the Debye screening mass is
$m_D^2=\frac{2N_{c,f}g^2}{T}\int\frac{d^3\k}{(2\pi)^3} n(k)[1\pm n(k)]
=\big(N_c+\frac{N_f}{2}\big)\frac{g^2T^2}{3}$ 
and HTL resummations are \cite{Braaten:1989mz,Weldon:1982aq}
\begin{eqnarray}
\Pi_L(Q)&=&m_D^2\left[1-\frac{\omega}{2q}\left(\ln
\frac{q+\omega}{q-\omega}-i\pi\right)\right] \, ,
\nonumber\\
\Pi_T(Q)&=&m_D^2\left[\frac{\omega^2}{2q^2}+\frac{\omega(q^2-\omega^2)}
{4q^3}\left(\ln\frac{q+\omega}{q-\omega}-i\pi\right)\right] \, .
\end{eqnarray}
In a leading-log approximation, we have \cite{Moore:2004tg} 
\begin{eqnarray}
\label{llog}
\kappa_L(p)&=&\kappa_0\frac{3}{2}
\left[\frac{E^2}{p^2}-\frac{E(E^2-p^2)}{2p^3}\ln\frac{E+p}{E-p}\right]
 \, ,
\nonumber\\
\kappa_T(p)&=&\kappa_0\frac{3}{2}
\left[\frac{3}{2}-\frac{E^2}{2p^2}+\frac{(E^2-p^2)^2}{4Ep^3}\ln\frac{E+p}{E-p}\right]
\, ,
\end{eqnarray}
where 
$\kappa_0\equiv\kappa_L(p=0)=\kappa_T(p=0)=\frac{g^4C_FT^3}{18\pi}\Big(N_c+\frac{N_f}{2}\Big)\left[\ln\frac{T}{m_D}+\mathcal{O}(1)\right]$.

As the heavy quark distribution must approach the thermal equilibrium, 
$f(\p)\propto e^{-E_{\p}/T}$, the drag coefficient and the longitudinal 
diffusion coefficient are related by $\eta(p)=\kappa_L(p)/(2TE)$ to 
leading order in $T/E$. 
At this order, the collisional energy loss of heavy quarks, 
$-\frac{dE}{dz}=p\eta(p)$, is also proportional to the longitudinal 
diffusion coefficient.

\section{Radiative Effects}
\label{radiation}

The collisional energy loss by diffusion is dominant for low-momentum 
heavy quarks, whereas the medium-induced gluon emission starts 
to contribute as the heavy quark momentum increases. 
Unlike quasiparticle dynamics where both collisional and radiative processes 
contribute at leading order \cite{Arnold:2002zm}, gluon emission off slow 
heavy quarks is $\mathcal{O}(g^2)$ suppressed than elastic scatterings at weak 
coupling. 
While there is $\mathcal{O}(1/g^2)$ enhancement for light partons with soft 
gluon exchange and collinear gluon emission 
\cite{Aurenche:1998nw,Arnold:2001ba}, 
radiation from heavy quarks depends on their momentum extent because the 
heavy quark mass cannot be ignored in an intermediate-momentum regime.  
At higher orders, diffusion and radiation are not clearly distinguished 
\cite{Ghiglieri:2015ala}. 
We will see a part of radiative effects contributes to the 
transverse momentum diffusion coefficient.

The radiative energy loss of ultrarelativistic partons, known as jet 
quenching, has been extensively studied using different formalisms: the 
path-integral formulation, a Schr\"{o}dinger-like equation, opacity and 
high-twist expansions, and a summation of ladder diagrams 
\cite{Zakharov:1996fv,Baier:1996kr,Gyulassy:2000er,Wang:2001ifa,Arnold:2002ja}. 
Gluon emission from light partons takes some time (called the formation 
time), $t_f\sim 1/(g^2T)$ which is of the same order as the mean free path. 
In that case, we need to sum multiple scatterings which reduce the emission 
rate due to the coherence (LPM) effect \cite{Landau:1953um,Migdal:1956tc}. 
The radiative energy loss of heavy quarks has been evaluated within the 
frameworks of high-twist and opacity expansions 
\cite{Zhang:2003wk,Djordjevic:2003zk,Djordjevic:2009cr,Armesto:2003jh,Abir:2015hta}. 
In this work, we will follow a diagrammatic approach of Ref. 
\cite{Arnold:2001ba} to evaluate gluon emission from heavy quarks with $p\gg m$.

\begin{figure}
\includegraphics[width=0.9\textwidth]{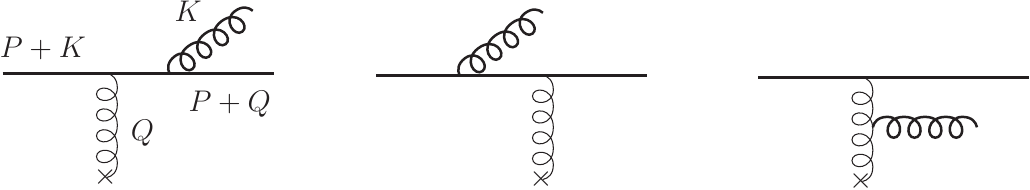}
\caption{
Gluon radiation off heavy quarks interacting with soft classical fields. 
Thick solid lines denote heavy quarks, thick and thin wiggly 
lines are hard($K\sim T$) and soft($Q\sim gT$) gluons, respectively, and 
crosses are for thermal scattering centers.
}
\label{gb}
\end{figure}

For energetic heavy quarks, soft collisions induce collinear 
gluon-bremsstrahlung.
Fig. \ref{gb} shows diagrams for the radiative contributions 
\cite{Gunion:1981qs}. 
The radiative energy loss is dominated by hard gluon emission ($k\sim T$), 
even though the energy of gluon is still much smaller than that of heavy quark 
($k\ll E_{\p})$. 
In the collinear limit, the emitted gluon has transverse momentum, 
$k_T\sim gT$. 
The radiative process is then factorized into elastic scattering and the 
gluon emission factor allowing enhancement so that radiation can be as 
important as elastic scattering.

The energy change in the radiation process is the inverse formation time,  
\begin{equation}
\frac{1}{t_f}=\delta E=E_{\p}+k^0-E_{\p+\k}
\simeq\frac{\k_T^2+m^2x^2+m_g^2}{2k(1-x)} \, ,
\end{equation}
where $x=k/E_{\p+\k}$ and $m_g^2=m_D^2/2$ is the thermal mass 
of the emitted gluon. 
We have chosen the initial transverse momentum of heavy quark to be zero, 
$\p_T+\k_T=0$. 
If the heavy quark momentum is so large that $mx\sim gT$, then 
we need to consider multiple soft scatterings as for light partons. 
On the other hand, radiation rarely occurs from heavy quarks 
with $p\lesssim m$ if $mx\sim T$. 
To smoothly interpolate between the two limits, we will consider only 
$gT\ll mx\ll T$ case. 
Then the formation time is shorter than the mean free path, allowing us to  
limit our discussion to gluon emission from a single scattering.

Gluon emission from quark-gluon plasmas has been computed by summing multiple 
scatterings during the emission process 
\cite{Arnold:2001ba,Arnold:2001ms,Arnold:2002ja}. 
Without the LPM effect, the radiative corrections to the collision kernel and 
the transport coefficient $\hat{q}$ have been evaluated for ultrarelativistic 
partons \cite{Liou:2013qya,Blaizot:2013vha,Ghiglieri:2022gyv},  
but not for heavy quarks with finite mass effects.
Adopting a similar approach to $\hat{q}$ in a single scattering, we will 
consider the heavy quark case in this work.     
In this way, heavy quark diffusion and radiation can be consistently 
calculated using the transverse momentum diffusion coefficient in an 
intermediate-momentum regime.

The gluon emission rate is given by \cite{Jeon:2003gi} 
\begin{equation}
\label{Grate}
\frac{d\Gamma(E_{\p},k)}{dk}=
\frac{g^2C_F}{8\pi k^3}[1+n_B(k)][1-n_F(E_{\p-\k})]
\frac{(1-x)^2+1}{(1-x)^2}\int \frac{d^2\p_T}{(2\pi)^2}\p_T\cdot{\rm Re} \,
 F(\p_T) \, ,
\end{equation}
where $\Gamma(E_{\p},k)$ is the rate for heavy quark with momentum $\p$ to 
emit a gluon with energy $k$, $n_B(k)$ and $n_F(E_{\p-\k})$ are the 
Bose-Einstein and Fermi-Dirac thermal distributions, respectively, and  
$F(\p_T)$ is the solution of a linear integral equation which sums 
ladder diagrams. 
For a single scattering (see appendix \ref{emit}), 
\begin{equation}
\label{sol}
{\rm Re} \, F(\p_T)=\frac{2}{\delta E(\p_T)}\int\frac{d^3\q}{(2\pi)^3}
C(\q)
\left[\frac{\p_T}{\delta E(\p_T)}-\frac{\p_T+\q_T}{\delta E(\p_T+\q_T)}\right] \, ,
\end{equation}
where $C(\q)$ is the collision kernel in Eq. (\ref{kernel}). 
Now, we take the real processes 
\cite{Ghiglieri:2013gia} and assume that the 
emitted gluon has a larger transverse momentum than the soft momentum of gluon 
exchange, $p_T\gg q_T$\footnote{The approximations and power-counting used 
in this section are similar to those for semi-collinear emission 
\cite{Ghiglieri:2013gia,Ghiglieri:2015zma,Ghiglieri:2015ala} or 
soft-collinear effective theory 
\cite{Abir:2015hta}.}: 
\begin{eqnarray}
\label{rad}
\int\frac{d^2\p_T}{(2\pi)^2}\p_T\cdot{\rm Re} \, F(\p_T)
&=&\int \frac{d^2\p_T}{(2\pi)^2}
\int \frac{d^3\q}{(2\pi)^3}C(\q)
\left[\frac{\p_T}{\delta E(\p_T)}-\frac{\p_T+\q_T}{
\delta E(\p_T+\q_T)}\right]^2 \, ,
\nonumber\\
&\simeq&8\kappa_{T}k^2(1-x)^2\int \frac{d^2\p_T}{(2\pi)^2}
\frac{1}{(\p_T^2+m^2x^2+m_g^2)^2} \, ,
\end{eqnarray}
where we have used the definition of the transverse momentum diffusion 
coefficient, Eq. (\ref{kappa}). 
Except for employing the collision kernel responsible for heavy quark 
diffusion instead of a static Debye-screened potential or the same kernel 
as light partons in a dynamical medium \cite{Aurenche:2002pd}, this 
corresponds to the incoherent limit of the $N=1$ opacity expansion 
\cite{Djordjevic:2003zk,Armesto:2003jh,Djordjevic:2009cr}.

In the Boltzmann equation Eq. (\ref{boltz}), the radiation term is given by 
\cite{Arnold:2002ja,Jeon:2003gi}
\begin{equation}
\label{C12}
C_{\rm rad}[f]\sim 
\int dk \left[
f(\p+\k)\frac{d\Gamma(E_{\p+\k},k)}{dk}
-f(\p)\frac{d\Gamma(E_{\p},k)}{dk}\right] \, ,
\end{equation}
where $\p+\k\simeq(p+k)\hat{\p}$ in the eikonal approximation. 
$k<0$ corresponds to gluon absorption which is required for detailed balance. 
Heavy quark radiation is different from light partons in that gluon 
emission is suppressed at smaller angles than $m/E$ 
\cite{Dokshitzer:2001zm}. 
This dead-cone effect can be observed if $m^2x^2$ is larger than the other 
terms in the denominator of Eq. (\ref{rad}).  
In the region $gT\ll mx\ll T$ of our interest, the radiation term can be larger 
than $\mathcal{O}(g^6)$ but smaller than $\mathcal{O}(g^4)$ of the 
ultrarelativistic limit.  
If the energy carried by an emitted gluon is soft ($k\sim gT$), we can expand 
the first term in Eq. (\ref{C12}), which contributes to the longitudinal 
diffusion at next-to-leading order $\mathcal{O}(g^5)$ 
\cite{Ghiglieri:2015ala}.

The collision kernel $C(k_T)$ is the rate for heavy quark to acquire 
transverse momentum $k_T$. 
After gluon emission in Fig. \ref{gb}, radiative corrections arise, 
\begin{equation}
\label{deltaC}
\delta C(k_T)=
\frac{g^2 C_F \kappa_{T}}{\pi}
\int\frac{dk}{k}[(1-x)^2+1]
\frac{1}{(\k_T^2+m^2x^2+m_g^2)^2} \, ,
\end{equation}
which has been obtained in the same approximation as Eq. (\ref{rad}). 
Then we have the radiative correction to the transverse momentum diffusion 
coefficient, 
\begin{equation}
\label{dkappaT}
\delta \kappa_T(p)
=\frac{1}{2}\int\frac{d^2\k_T}{(2\pi)^2}k_T^2\delta C(k_T)[1+n_B(k)] \, ,
\end{equation}
Using the kinematic boundaries,  
$k_{T,\rm max}\sim k$, $k_{\rm max}\sim p$, and $k_{\rm min}\sim T$, 
\begin{equation}
\delta\kappa_T(p)\sim g^2\kappa_{T}\ln\frac{E}{m}
\ln\frac{p}{T} \, .
\end{equation}
In comparison to the leading-order coefficient $\kappa_T$, 
$\delta \kappa_T$ is $\mathcal{O}(g^2)$ 
suppressed but logarithmically enhanced in the high-momentum limit.  
This is analogous to quantum corrections to the transverse momentum 
broadening coefficient \cite{Liou:2013qya,Blaizot:2013vha}, except for the 
different phase space boundaries and the heavy quark mass regulating the 
collinear singularity.   
The importance of the factor $[1+n_B(k)]$ in Eq. (\ref{dkappaT}) has been 
discussed in Ref. \cite{Ghiglieri:2022gyv}: it is needed to account for 
Bose-enhancement for $k\lesssim T$, connecting to $\mathcal{O}(g)$ classical 
corrections for soft-gluon emission. 
A numerical estimate for this potentially large correction is given in 
Fig. \ref{kappaLT} (a) in the next section. 
The correction increases with the heavy quark momentum and becomes comparable 
to the leading-order coefficient at high momentum.

The final form of the radiation term is given by  
\begin{equation}
\label{Crad}
C_{\rm rad}[f]= 
\int dk \left[
f((p+k)\hat{\p})\frac{d\Gamma(E_{(p+k)\hat{\p}},k)}{dk}
-f(\p)\frac{d\Gamma(E_{\p},k)}{dk}\right]
+\frac{1}{2}\nabla_{\p_T}^2[\delta\kappa_T(p) f(\p)] \, .
\end{equation}
Because the emission rate in Eq. (\ref{C12}) can be as small as 
$\mathcal{O}(g^6)$ at low momentum, we have included the radiative correction 
($\delta\kappa_T$ term) to the eikonal approximation.

\section{Numerical Analysis}
\label{numerical}

We have formulated the heavy quark Boltzmann equation with diffusion and 
radiation in Eqs. (\ref{Ccol}) and (\ref{Crad}), respectively. 
Using the leading-log momentum dependence, Eq. (\ref{llog}), 
the two collision terms involve only one parameter, the static momentum 
diffusion coefficient, $\kappa_{L,T}(p=0)\equiv\kappa_0=2T^2/D_s$ ($D_s$ is 
the spatial diffusion coefficient at $p=0$). 
Since the perturbative expansion poorly converges at a realistic value of the 
strong coupling constant \cite{Caron-Huot:2007rwy}, we will use $\kappa_0$ 
constrained by lattice QCD data so that nonperturbative effects can be 
absorbed in the transport coefficient. 
Employing $\kappa_0$ in this way amounts to effectively changing the coupling 
constant and the thermal masses of light partons in the collision kernel,  
Eq. (\ref{kernel}).

\begin{figure}
\includegraphics[width=0.45\textwidth]{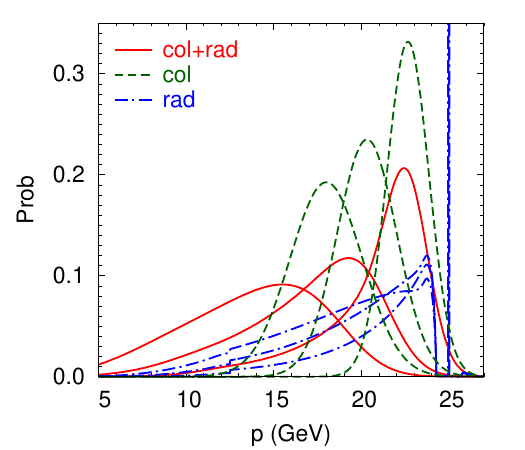}
\caption{
The probability distribution of $b$ quarks with initial momentum $p_0=25$ GeV 
in a static medium at $T=300$ MeV. 
From right to left, $t=5, \, 10, \, \mbox{and} \, 15$ fm. 
We have used $m=4.5$ GeV, $(2\pi T)D_s=6$, and $\alpha_s=0.3$ for gluon 
emission. 
}
\label{delta}
\end{figure}

Figure \ref{delta} shows how the $b$ quark distribution with an initial 
delta function evolves in a static medium, under the influence of two 
different types of energy loss. 
We notice that how the distributions are spread out with time depends on the 
energy loss mechanism.  
The diffusion process is characterized by Gaussian fluctuations,   
whereas the radiative one develops non-Gaussian distributions.  
It has been discussed that there are significant differences between Langevin 
and Boltzmann approach for heavy quark diffusion unless the 
ratio $m/T$ is large: the Langevin(Fokker-Planck) approach is a good 
approximation for bottom quark diffusion \cite{Das:2013kea}. 
In our formulation, the radiation term of Eq. (\ref{C12}) is not expanded 
for soft gluon emission, so it is not a diffusion operator. 
This difference between diffusion and radiation might allow the medium 
modifications by two mechanisms to be qualitatively distinguishable from 
each other.

\begin{figure}
\includegraphics[width=0.45\textwidth]{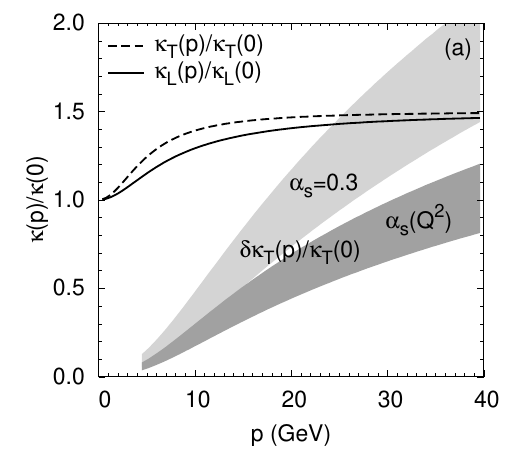}
\includegraphics[width=0.45\textwidth]{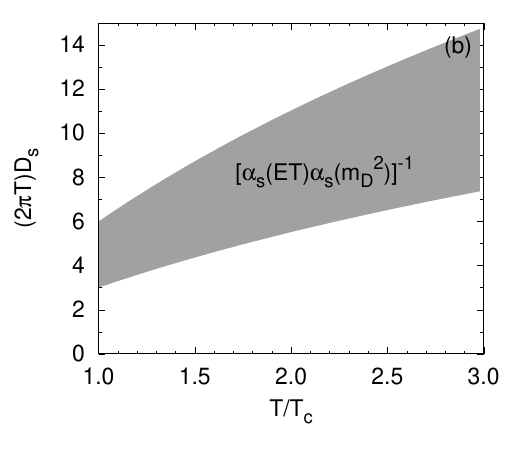}
\caption{
(a) The momentum dependence of the heavy quark transport coefficients. 
The light and dark shaded regions represent the momentum-dependent 
$\delta\kappa_T(p)$ using fixed and running coupling constants, 
respectively.  
The upper(lower) lines of the shaded regions correspond to $T=157(475)$ MeV. 
(b) The temperature dependence of $(2\pi T)D_s$. 
}
\label{kappaLT}
\end{figure}

The transport coefficients and their dependence on momentum and temperature 
are crucial to analyze experimental data. 
Fig. \ref{kappaLT} shows the momentum and temperature dependence of the 
transport coefficients employed in this work.  
As the momentum of heavy quark increases, the momentum diffusion coefficient 
and energy loss increase. 
At the leading-log order, the momentum dependence of the longitudinal and 
transverse diffusion coefficients is modest, shown as the solid and dashed 
lines, respectively. 
As mentioned in the previous section, $\delta\kappa_T(p)$ (the radiative 
correction to $\kappa_T$) also grows with momentum and becomes 
considerable at high momentum, especially for a strong coupling constant 
$\alpha_s\sim 0.3$.

The temperature dependence of $(2\pi T)D_s$ comes from running 
of the coupling constant\footnote{The running coupling constant is related 
to nonperturbative effects in heavy quark diffusion. 
These effects have also been considered in the T-matrix approach 
\cite{He:2011qa} and using a rather strong coupling with large quasiparticle 
masses near $T_c$ \cite{Song:2015ykw,Scardina:2017ipo}.}. 
An infrared-finite effective running coupling has been developed and employed 
for the spacelike momentum transfer \cite{Dokshitzer:1995qm,Gossiaux:2008jv}. 
Replacing the coupling constant in the t-channel amplitude by the 
running coupling and using the one-loop result,
\begin{equation} 
\alpha_s(Q^2)=
\frac{12\pi}{\big(11N_c-2N_f)\ln(Q^2/\Lambda_{\rm QCD}^2)} \, , 
\end{equation}
at the scale $Q^2\sim$ t ($\Lambda_{\rm QCD}\approx 200$ MeV), resummations 
and nonperturbative effects can be implemented 
\cite{Dokshitzer:1995qm,Peshier:2006hi,Gossiaux:2008jv}.  
In this work, we follow Ref. \cite{Peigne:2008nd} to consider dependence on 
a wide range of t scales, from $\mathcal{O}(m_D^2)$ up to $\mathcal{O}(ET)$. 
Then $\kappa_0\propto\alpha_s(ET)\alpha_s(m_D^2) \, T^3$, where $m_D$ is 
self-consistently determined by \cite{Peshier:2006ah}
\begin{equation}
\ln\left(\frac{m_D^2}{\Lambda_{\rm QCD}^2}\right)
=\frac{N_c(1+N_f/6)}{11N_c-2N_f}\left(\frac{4\pi T}{m_D}\right)^2 \, .
\end{equation}
As the temperature decreases, the running coupling becomes stronger near 
$T_c$ where nonperturbative effects enter. 
For temperature and momentum considered in this work, 
$\alpha_s\sim 0.23-0.68$ which is of the same order as the 
effective coupling from Ref. \cite{Gossiaux:2008jv}. 
While the coupling constant decreases with increasing temperature, 
$(2\pi T)D_s=4\pi T^3/\kappa_0\propto [\alpha_s(ET)\alpha_s(m_D^2)]^{-1}$ 
increases by a factor of $\sim 2.5$ in Fig. \ref{kappaLT} (b), aligning 
closely with the lattice QCD data from Refs. 
\cite{Banerjee:2011ra,Francis:2015daa,Banerjee:2022gen}. 
Although the degree of increase might vary with a different choice of effective 
coupling, the temperature dependence is expected to be qualitatively 
consistent with that in the current study.  
For the radiation process, we follow Ref. \cite{Djordjevic:2013xoa} to 
determine the running coupling constant at the scale 
$Q^2=(\k_T^2+m^2x^2+m_g^2)/x$.

The nuclear modification factor of heavy mesons is an important observable to 
measure the thermal medium effects in heavy-ion collisions. 
It is affected by the initial production of heavy quarks, medium evolution, 
and hadronization as well as heavy quark interactions in quark-gluon 
plasmas. 
In this work, we focus on the energy loss effects in quark-gluon plasmas, 
especially the qualitative difference between two energy loss mechanisms. 
To isolate significant uncertainties related to medium expansion and  
hadronization, we assume a simple model. 
For the initial spectrum of $b$ quarks, we take the differential cross 
section of $B$ meson production measured in $pp$ collisions 
\cite{CMS:2017uoy}, fit to the following form: 
\begin{equation}
\frac{dN}{p_Tdp_T}\propto \frac{1}{(\p_T^2+\Lambda^2)^{\alpha}} \, ,
\end{equation} 
where $\Lambda=6.07$ GeV and $\alpha=2.85$. 
Then the plasma evolution is described by a Bjorken expansion, 
$T(t)=T_0(t_0/t)^{1/3}$ \cite{Bjorken:1982qr} with $t_0=0.6$ fm and 
$T_0=475$ MeV \cite{Alberico:2013bza} until 
$T_c=157$ MeV \cite{HotQCD:2018pds}. 
These initial conditions depend on centrality and collision energy, but the 
variations of the values have little impact on our qualitative analysis of 
the momentum spectrum. 
After solving for the heavy quark distribution, we take the ratio of the final 
spectrum to the initial one to estimate the suppression factor, 
\begin{equation}
R_{AA}(p_T)= 
\frac{\, \frac{dN}{dp_T}\Big\vert_{t=t_f} \, }
{ \, \frac{dN}{dp_T}\Big\vert_{t=t_0} \, } \, .
\end{equation}

\begin{figure}
\includegraphics[width=0.45\textwidth]{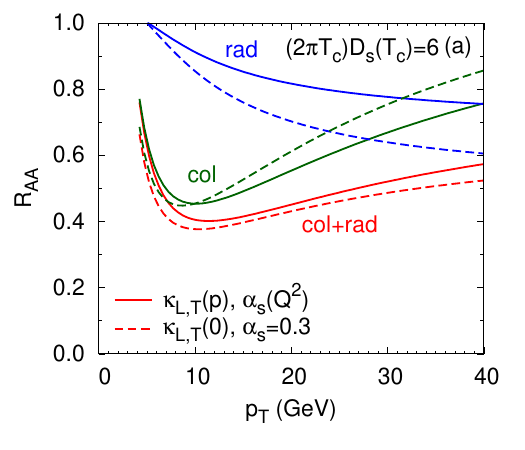}
\includegraphics[width=0.45\textwidth]{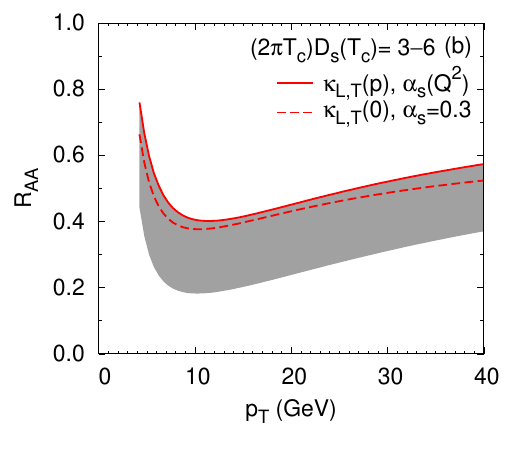}
\caption{
The nuclear modification factor $R_{AA}$ for $b$ quarks. 
(a) The solid lines show the results using the momentum-dependent 
$\kappa_{L,T}$ and running $\alpha_s$, with $(2\pi T_c)D_s(T_c)=6$ fixed at 
$p=0$ and $T=T_c$.   
The dashed lines show the results with constant $\kappa_{L,T}$ and $\alpha_s$. 
(b) The upper and lower bounds of the shaded region correspond to 
$(2\pi T_c)D_s(T_c)=6$ and $3$, respectively. 
}
\label{raa}
\end{figure}

Figure \ref{raa} shows the nuclear modification factor for $b$ quarks. 
The solid lines are the results using the momentum-dependent diffusion 
coefficients and the running coupling constant, while the dashed lines are the 
results with constant diffusion coefficient and coupling constant. 
At $p=0$ and $T=T_c$, we have fixed $(2\pi T_c)D_s(T_c)=3-6$, closely aligning 
with the lattice QCD data from Refs. 
\cite{Banerjee:2011ra,Francis:2015daa,Banerjee:2022gen}. 
The value of $\alpha_s$ directly affects the suppression by the radiative 
energy loss: the stronger the coupling, the smaller the $R_{AA}$ factor. 
As expected from Fig. \ref{delta}, the collisional and radiative effects 
exhibit distinct momentum behaviors. 
The $R_{AA}$ by the radiative energy loss consistently decreases with momentum, 
while the $R_{AA}$ by the collisional energy loss decreases at low momentum 
but increases at intermediate momentum. 
Thus, as the heavy quark momentum rises, the dominant energy loss shifts 
from collisional to radiative. 
We note that the momentum at which this transition occurs depends 
on the transport coefficients and their dependence on momentum and 
temperature.   
In our numerical analysis, the transition takes place (and radiation becomes 
effective) at higher momentum 
when $\kappa_{L,T}$ increases with momentum and $\alpha_s$ decreases with 
energy and temperature, compared to when they are constant.

While the momentum-dependence of Eq. (\ref{llog}) is valid to leading 
logarithm in $T/m_D$, higher-order terms can influence the flatness of 
the suppression factor. 
To estimate this effect, if we consider a $30\%$ increase in the diffusion 
coefficients' growth rate with respect to momentum, the $R_{AA}$ factor with 
$(2\pi T_c)D_s(T_c)=6$ would be reduced by at most $20\%$ at high momentum, 
flattening $R_{AA}$. 
Despite the stronger momentum-dependence, it would still be within the shaded 
region in Fig. \ref{raa} (b) due to the large uncertainties of $D_s$. 
The qualitative behavior discussed in the previous paragraph remains 
consistent because the momentum-dependence enters both diffusion and 
radiation simultaneously.     
This phenomenological study estimates the suppression factor with the leading 
momentum-dependence of the diffusion coefficients, allowing for implicit 
inclusion of higher-order effects through the nonperturbative lattice QCD 
data and the running coupling constant.

The qualitative distinction between diffusion and radiation in the momentum 
spectra might be useful to identify the relevant energy loss 
process\footnote{To discriminate between the collisional and radiative 
energy loss mechanisms, angular correlations of heavy quark pairs have also 
been studied \cite{Nahrgang:2013saa,Cao:2015cba}.}.
The radiative effect makes the nuclear modification factor flatter than the 
suppression entirely by the collisional one,  
as seen in Fig. \ref{raa} (a).    
Although it is premature to compare our numerical results with experimental 
data, the suppression factor calculated with $(2\pi T_c)D_s(T_c)=3-6$ is 
comparable with the $R_{AA}$ factor of $B$ mesons 
\cite{CMS:2017uoy,CMS:QM2023}. 
A Bjorken expansion has been employed in this work, while 
(3+1)-dimensional expansion provides the time evolution of 
the spatial distribution of temperature and collective flow velocity. 
The energy loss of heavy quark will be influenced by a modified profile of 
quark-gluon plasmas, determined by different temperature, lifetime, and 
expansion rate of (3+1)-dimensional evolution. 
However, similar medium modifications, averaged 
over position, are expected through the adjustment of $D_s$.  
In future work, we plan to perform a more quantitative analysis with 
realistic hydrodynamic evolution and hadronic effects.

We mention that the valid momentum range, where gluon emission from a single 
scattering is applicable, is not clear. 
In a high-momentum regime, the emission rate must be computed in multiple 
soft scatterings.   
Although gluon emission is more involved than photon emission (because gluons 
carry color) \cite{Arnold:2002ja}, the LPM effect on the photon emission rate 
for $k\gtrsim 2T$ is less than $30\%$ \cite{Arnold:2001ms}. 
If we include this suppression in our radiation term, the $R_{AA}$ 
factor is expected to increase slightly with momentum, 
approximately $\sim10\%$ at most.  
However, the momentum dependence of the heavy quark spectrum does not change 
significantly. 
We still expect to differentiate the radiative contribution from the diffusion 
effects in an intermediate-momentum regime.

\begin{figure}
\includegraphics[width=0.45\textwidth]{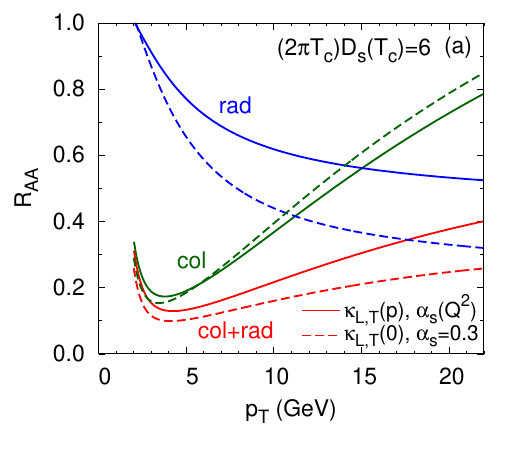}
\includegraphics[width=0.45\textwidth]{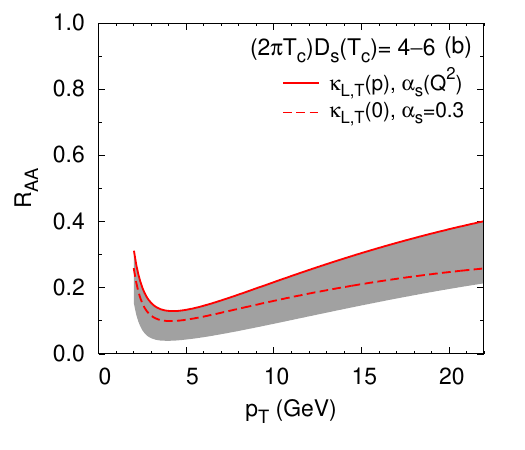}
\caption{
The estimated $R_{AA}$ factor for $c$ quarks, with $m=1.5$ GeV and 
the initial spectrum given by the differential cross 
section of $D$ meson \cite{ALICE:2019nxm}. 
}
\label{charm_raa}
\end{figure}

Compared to bottom quarks, charm quarks have $3$ times smaller mass, thus 
the energy loss is expected to be larger.  
Although the heavy quark conditions and approximations assumed in our model 
may be only marginally satisfied for charm quarks, we have applied our 
formulation to demonstrate the impact of the heavy quark mass (see Fig. 5).  
Charm quarks are more suppressed by elastic scattering and 
gluon-bremsstrahlung than bottom quarks, while the $R_{AA}$ factor depends 
similarly on momentum and temperature through the transport coefficients. 
The transition between diffusion and radiation occurs at relatively lower 
momentum, and thus the radiative effects become more significant to determine 
the intermediate-momentum spectrum.

\section{Summary}
\label{summary}

In this work, we have formulated the heavy-quark Boltzmann equation with 
diffusion and radiation from a single scattering in an intermediate-momentum 
regime.  
As a part of the radiative effects, we have obtained quantum corrections 
to the transverse momentum diffusion coefficient, which are $\mathcal{O}(g^2)$ 
suppressed than the leading-order diffusion coefficient but logarithmically 
enhanced in the high-energy limit. 
Employing the same collision kernel consistently for both 
processes, our formulation has only a single transport parameter, the static 
diffusion coefficient which can be constrained by nonperturbative 
determination.  
Although our approach is based on perturbation, the running coupling constant 
and the diffusion coefficient given by lattice QCD data allow for 
nonperturbative effects at low momentum and temperature.

We have investigated the momentum dependence of the heavy quark spectrum and 
the suppression factor, determined by the two types of heavy quark 
energy loss. 
For nearly collinear gluon emission from a single scattering, the medium 
modifications by radiation are found to be distinguishable from those by 
diffusion so that the relevant energy loss mechanism can be identified.   
Our numerical results indicate that, at low and high momentum, 
the $R_{AA}$ factor is primarily influenced by the collisional and radiative 
energy loss, respectively. 
Meanwhile, the importance of the radiative effects at intermediate 
momentum is determined by the momentum-dependent diffusion coefficient and 
the running coupling constant.

We have concentrated on the qualitative features of the heavy quark momentum 
spectra in quark-gluon plasmas. 
Eventually to describe the experimental data of heavy mesons, we need to 
consider other effects such as hadronization, finite-size medium, viscous 
corrections in the hydrodynamic expansion 
\cite{Sarkar:2018erq,Kurian:2020orp,Singh:2023smw}, 
and possible pre-equilibrium dynamics 
\cite{Boguslavski:2023fdm,Boguslavski:2023alu}. 
In the same framework, it is also essential to describe the elliptic flow 
induced by the spatial anisotropy of thermal media. 
Although various transport models for heavy quarks have been developed, 
incorporating both elastic and inelastic scatterings 
\cite{Cao:2016gvr,Cao:2017crw,Zigic:2018ovr,Ke:2018tsh}, 
there still exist large uncertainties in an intermediate-momentum regime.  
We hope that our approach provides a way to understand the transition between 
diffusion and radiation and to distinguish the radiative effects in the heavy 
quark momentum spectra.

\appendix

\section{Gluon Emission}
\label{emit}

\begin{figure}
\includegraphics[width=0.4\textwidth]{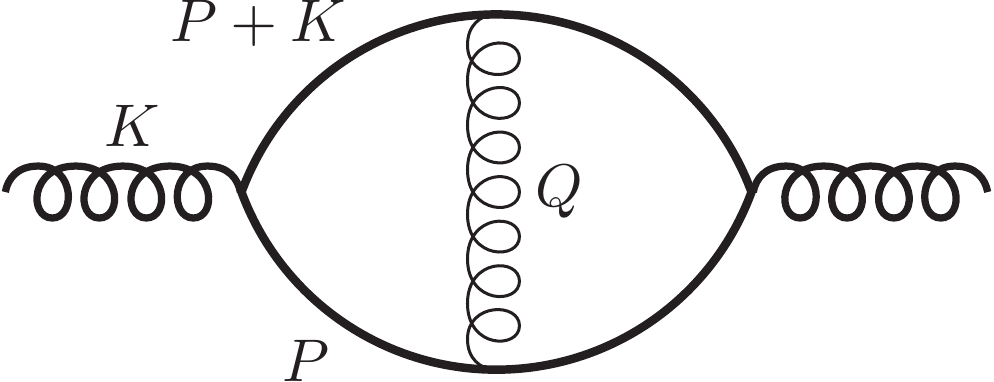}
\caption{
Gluon emission from a single scattering. 
$P$ and $K$ are nearly collinear ($k_T\sim gT$) and 
the gluon exchange is soft ($Q\sim gT$).  
}
\label{loop}
\end{figure}

In the high-momentum limit, gluon emission from heavy quarks is 
akin to that from light partons involving multiple scatterings. 
For the rigorous derivation of an integral equation which sums multiple 
scatterings, we refer to Refs. \cite{Arnold:2001ba,Arnold:2002ja}. 
In this appendix, we use the same approach to evaluate a single 
gluon exchange diagram, Fig. \ref{loop} which is relevant to the radiative 
energy loss of heavy quarks. 
Although the emitted gluon can also interact with soft background fields, 
the emission rate can be simplified by assuming the real processes with 
$k_T\gg q_T$ as in Section \ref{radiation}.

A heavy quark loop in ladder diagrams involves the following frequency 
integral:  
\begin{equation}
\label{pinching}
\int\frac{dp^0}{2\pi}
\frac{1}{p^0-E_{\p}+i\Gamma/2} \,
\frac{1}{p^0+k^0-E_{\p+\k}-i\Gamma/2}
\simeq\frac{1}{i\delta E+\Gamma} \, ,
\end{equation}
where $\Gamma/2$ is the heavy quark damping rate 
\cite{Pisarski:1993rf}. 
In the ultrarelativistic limit ($\delta E\sim g^2T$), this allows 
$\mathcal{O}(1/g^2)$ enhancement so that gluon-bremsstrahlung contributes at 
leading order.

In the kinematic regime with $t_f\ll 1/(g^2T)$, 
soft gluon exchange is perturbation.    
Based on a Bethe-Salpeter equation for the gluon vertex from either side of 
the diagram, Fig. \ref{loop} is roughly expressed as the sum of the loop 
diagrams without and with a single gluon exchange,  
\begin{equation}
F(\p_T)=\frac{2\p_T}{i\delta E+\Gamma}+\frac{1}{i\delta E+\Gamma}
\int\frac{d^3\q}{(2\pi)^3}C(\q)F(\p_T+\q_T) \, ,
\end{equation}
where $\p_T$ is the transverse projection with respect to $\k$, and $C(\q)$ is 
the collision kernel of Eq. (\ref{kernel}). 
Then, multiplying both sides by $i\delta E+\Gamma$ and using 
$\Gamma=\int\frac{d^3\q}{(2\pi)^3}C(\q)$, we obtain the following 
integral equation:
\begin{equation}
\label{inteq}
2\p_T=i\delta E \, F(\p_T)
+\int\frac{d^3\q}{(2\pi)^3}C(\q)\left[F(\p_T)-F(\p_T+\q_T)\right] \, .
\end{equation}
Since $\delta E$ is larger than $\int\frac{d^3\q}{(2\pi)^3}C(\q)\sim g^2T$, 
we can solve it perturbatively. 
The leading-order solution is pure imaginary, 
$F^0(\p_T)=2\p_T/(i\delta E)$. 
Substituting this into the equation, we obtain the next-order 
whose real part determines the emission rate in Eqs. (\ref{Grate}) and
 (\ref{sol}).

\section*{Acknowledgments}

I would like to thank Sangyong Jeon, Che-Ming Ko, Su Houng Lee, Peter Levai, 
and Ralf Rapp for useful discussions and comments.  
This research was supported by Basic Science Research Program through the
National Research Foundation of Korea (NRF) funded by the Ministry of Education 
(No. 2021R1I1A1A01054927).

\end{document}